\documentclass[a4paper]{article}

\usepackage{INTERSPEECH2020}

\usepackage{siunitx}
\usepackage{subcaption}
\usepackage{graphicx}
\usepackage{hyperref}
\usepackage{url}
\usepackage[keeplastbox]{flushend}

\usepackage{localmath}

\title{Generalized Minimal Distortion Principle for Blind Source Separation}
\name{Robin Scheibler}
\address{LINE Corporation, Japan}
\email{robin.scheibler@linecorp.com}

\graphicspath{{figures/}{figures/20200507-012736_experiment2_config_sim_102af93240_plot_58c722c373/}{figures/20200511-112906_experiment1_config_sim_102af93240_plot_58c722c373/}}

\begin{document}

\maketitle
\begin{abstract}
  We revisit the source image estimation problem from blind source separation (BSS).
  We generalize the traditional minimum distortion principle to maximum likelihood estimation with a model for the residual spectrograms.
  Because residual spectrograms typically contain other sources, we propose to use a mixed-norm model that lets us finely tune sparsity in time and frequency.
  We propose to carry out the minimization of the mixed-norm via majorization-minimization optimization, leading to an iteratively reweighted least-squares algorithm.
  The algorithm balances well efficiency and ease of implementation.
  We assess the performance of the proposed method as applied to two well-known determined BSS and one joint BSS-dereverberation algorithms.
  We find out that it is possible to tune the parameters to improve separation by up to \SI{2}{\decibel}, with no increase in distortion, and at little computational cost.
  The method thus provides a cheap and easy way to boost the performance of blind source separation.
\end{abstract}

\section{Introduction}

Blind source separation (BSS) conveniently allows to separate a mixture signal into its constitutional components without the need for prior information~\cite{Comon:1512057}.
A pioneering method of BSS is independent component analysis (ICA) that only requires that the signal components be statistically independent and non-Gaussian~\cite{Comon:1994kr}.
In its canonical form, ICA tackles determined linear mixtures where the number of components is the same as that of sensors.
In this case, the separation task boils down to finding a square demixing matrix making the output components independent.
The determined BSS problem suffers from two inherent ambiguities.
First, any permutation of the sources in the output is equally acceptable.
Second, the sources may be scaled arbitrarily.
The first problem is particularly problematic in frequency-domain BSS (FD-BSS)~\cite{Smaragdis:1998kl}, where sources extracted at each frequency must be aligned.
This can be done via clustering~\cite{Sawada:fk}, or by considering the joint distribution over frequencies as in independent vector analysis (IVA)~\cite{Kim:2006ex,Hiroe:2006ib}.


Relatively less attention has been given to the scale ambiguity problem.
For FD-BSS on acoustic mixtures, the ambiguity is equivalent to an arbitrary filtering of sources.
Without a correction step, separated sources typically do not sound natural at all.
This can be addressed by estimating \textit{source images}, that is the source signal as perceived at the microphone locations.
There has been traditionally two ways of doing it.
First, the so-called \textit{projection back} (PB) method makes use of the linearity of determined BSS~\cite{Murata:2001gb}.
It relies on the observation that the columns of the inverse of the demixing matrix are \textit{steering vectors} for the sources.
This method may be unstable if the demixing matrix is poorly conditioned, which is not frequent, but may happen for some algorithms.
The second method applies the \textit{minimal distortion principle} (MDP) to adjust the scale compared to the input microphone signal.
Exploiting the independence of the other signals, it finds the filter minimizing the squared distance between the separated source and the input signal.
From a maximum likelihood point of view, this method assumes a Gaussian distribution of the residual when computing the distortion.
However, the residual is in fact the sum of the other sources and background noise.
Due to the non-Gaussianity of sources, it is unlikely to be Gaussian, leading to a sub-optimal choice for the scaling filter.
For example, residual sources may have some very large components.
A squared norm will try to reduce them, possibly at the cost of the target source.
For a detailed comparison and analysis of both methods, see~\cite{Koldovsky:2017ez}.

In this paper, we propose the generalized minimal distortion principle (GMDP) that uses the maximum likelihood estimator (MLE) for the image sources.
We futher propose to instantiate GMDP based on sparsity promoting mixed norms.
The intuition for using such a measure of distortion is that we want to allow the residual from minimal distortion to have some large entries.
The use of $\ell_p$-norms, and the $\ell_1$-norm in particular, for this purpose has been popularized by the LASSO algorithm~\cite{Tibshirani:1996dm} and the compressed sensing literature~\cite{Candes:2008hb}.
Another way to understand the use of such norms is via a generative model of the residual and maximum likelihood estimation.
For example, the $\ell_1$ norm corresponds to a Laplace model.
What we propose is to penalize the residual between the separated source and the reference input signal using a mixed norm $\ell_{p,q}$, for $0 < p \leq q \leq 2$.
The mixed norm allows to promote sparsity at different rates across time and mixtures (i.e. time and frequency for audio signals).
Unlike the $\ell_2$-norm, there is no closed-form solution for the mixed norm minimization.
Instead we rely on \textit{majorization-minimization} (MM) whereas a surrogate function dominating the objective is repeatedly minimized~\cite{Lange:2016wp}.
We construct the surrogate function from an inequality previously used in the context of sound field decomposition~\cite{Murata:2018ft}.
The final algorithm falls in the family of iteratively reweighted least-squares (IRLS), that has been heavily investigated in the context of sparse regression~\cite{Daubechies:2010hf,Kowalski:2009ge}.
We validate the proposed GMDP via large numerical simulations of determined speech separation.
We investigate the performance for several BSS algorithms: AuxIVA~\cite{Ono:2011tn}, ILRMA~\cite{Kitamura:2016vj}, and joint BSS and dereverberation ILRMA-T~\cite{Ikeshita:2019bl}.
We sweep values of $0 < p \leq q \leq 2$ for different number of sources and find that our approach outperforms both MDP and PB in terms of standard BSS metrics.
The code for the experiments is shared at \url{https://github.com/fakufaku/2020_interspeech_gdmp}.

The rest of this paper is as follows.
\sref{background} covers the conventional BSS scaling strategies.
\sref{gmdp} describes the proposed method.
The result of numerical experiments is shown in \sref{experiments}.

\section{Background}
\seclabel{background}

The notation in this paper is as follows.
We use lower and upper case bold letters for vectors and matrices, respectively.
Furthermore, $\mA^\top$ and $\mA^\H$ denote the transpose and conjugate transpose of matrix A, respectively.
The Euclidean norm of vector $\vv$ is $\|\vv\| = (\vv^\H \vv)^{1/2}$.
The diagonal matrix with $\vv$ on its diagonal is denoted $\diag(\vv)$.
Unless specified otherwise, indices k, m, f, and n are for source, sensor, frequency, and time, respectively.
They always take the ranges from the corresponding capital letter, i.e., $K$, $M$, $F$, and $N$, respectively.

We consider FD-BSS with $K$ sources and $M$ sensors in the short time Fourier transform (STFT) domain~\cite{Allen:1977in}.
The sensor inputs are described by the linear mixing model
\begin{equation}
  x_{mfn} = \sum_{k=1}^K h_{mkf} s_{kfn} + b_{mfn},
  \elabel{mix}
\end{equation}
where $x_{mfn},s_{kfn}\in \C$ are the $m$th sensor and $k$th source signals, respectively at frequency $f$ and frame $n$, and $h_{mkf}\in\C$ is the transfer function between the two.
The term $b_{mfn}$ optionally encompasses extra background noise and model mismatch.
We can conveniently group the sensor signals in the vector $\vx_{fn} = [x_{1fn},\ldots,x_{Mfn}]^\top$, and the sources similarly in $\vs_{fn}$ and $\vb_{fn}$, respectively.
Defining the channel matrix as $\mH_f \in \C^{M\times K}$ such that $(\mH_f)_{mk} = h_{mkf}$, \eref{mix} can be written in the compact form
\begin{equation}
  \vx_{fn} = \mH_f \vs_{fn} + \vb_{fn},\quad \forall f, n.
\end{equation}
Under this model, BSS algorithms may at best attempt to recover a source vector estimate $\vy_{fn}$ such that there is a mixing matrix $\mA_f\in\C^{M\times K}$ and
\begin{equation}
  \vx_{fn} \approx \mA_f \vy_{fn},\quad \forall f,n.
  \elabel{newmix}
\end{equation}
The scale ambiguity is clear since for any non-singular diagonal matrix $\mD$, $\mA_f \mD^{-1}$ and $\mD \vy_{fn}$ form an equally valid solution.
To avoid this scaling ambiguity, the \textit{source images} are sought instead. 
These are the sources as measured at a sensor location, e.g. the $k$th source at the $m$th sensor is $\hat{s}_{mkfn} = h_{mkf} s_{kfn}$.
When $\mA_f$ is available, then from its $k$th column $\va_{kf}$, the source images of the $k$th source can be obtained
\begin{equation}
  \hat{\vy}_{kfn} = \va_{fn} y_{kfn},
\end{equation}
so that $\vx_{fn} \approx \sum_k \hat{\vy}_{fn}$. Unfortunately, $\mA_f$ is typically not known and must also be estimated.


\subsection{Projection Back}

The so-called projection back technique is applicable when the number of sources is the same as that of sensors, i.e. $M=K$,~\cite{Murata:2001gb}, and $b_{mfn} = 0$ in \eref{mix}.
In this case, the demixing is typically done by estimating a square \textit{demixing matrix} $\mW_f \in \C^{M\times M}$ so that
\begin{equation}
  \vy_{fn} = \mW_f \vx_{fn},\quad \forall f,n.
\end{equation}
In that case, it is clear that \eref{newmix} holds with equality with $\mA_f = \mW^{-1}_f$.
Thus, the estimated source image is
\begin{equation}
  \hat{\vy}_{fn} = \mA_f \ve_k y_{kfn} = \mW_f^{-1} \ve_k y_{kfn}.
\end{equation}
where $\ve_k$ is the $i$th column of the identity matrix.
This method is widely used in practice and works reasonably well, as long as $\mW_f$ is well-conditioned.
This is usually the case as most BSS algorithms either impose its orthogonality, e.g.,~\cite{Hyvarinen:ek}, or penalize it with a log-determinant term, e.g.,~\cite{Ono:2011tn}.

\subsection{Minimal Distortion Principle}

In contrast, MDP finds the mixing weights that minimize the sum of squared differences between the separated source and the microphone inputs \cite{Matsuoka:2001da,Matsuoka:2002jy},
i.e., $\hat{\vy}_{fn} = \va_{kf} y_{kfn}$, with
\begin{equation}
  \va_{kf} = \underset{\va \in \C^M}{\arg\min}\ \E \| \vx_{fn} - \va y_{kfn} \|^2.
  \elabel{mdp}
\end{equation}
While derived originally from a different perspective, this source image estimator is optimal in the maximum likelihood sense when sources are uncorrelated and the background noise is Gaussian.
Under the uncorrelation assumption, one can show,
\begin{equation}
  \E\| \vx_{fn} - \va y_{kfn} \|^2 = \E \| \vh_{kf} s_{kfn} - \va y_{kfn} \|^2 + \text{const.},
\end{equation}
where $\vh_{kf}$ is the $k$th column of $\mH_f$.
Thus, \eref{mdp} is indeed the MLE of $\va_{kf}$ if $b_{mfn}$ is Gaussian.
The MDP can be shown to be equivalent to PB under the uncorrelation assumption~\cite{Koldovsky:2017ez}.
However, it is more stable and can deal with $K\neq M$.
In practice, however, both assumptions for its optimality are routinely violated.
In multivariate source models, such as IVA, uncorrelation is not required at all frequencies.
In addition, the background is typically not Gaussian.
We address these limitations in the next section.

\section{Generalized Minimal Distortion Principle}
\seclabel{gmdp}

\begin{figure*}
  \begin{subfigure}{.5\textwidth}
    \centering
    \includegraphics[width=\linewidth]{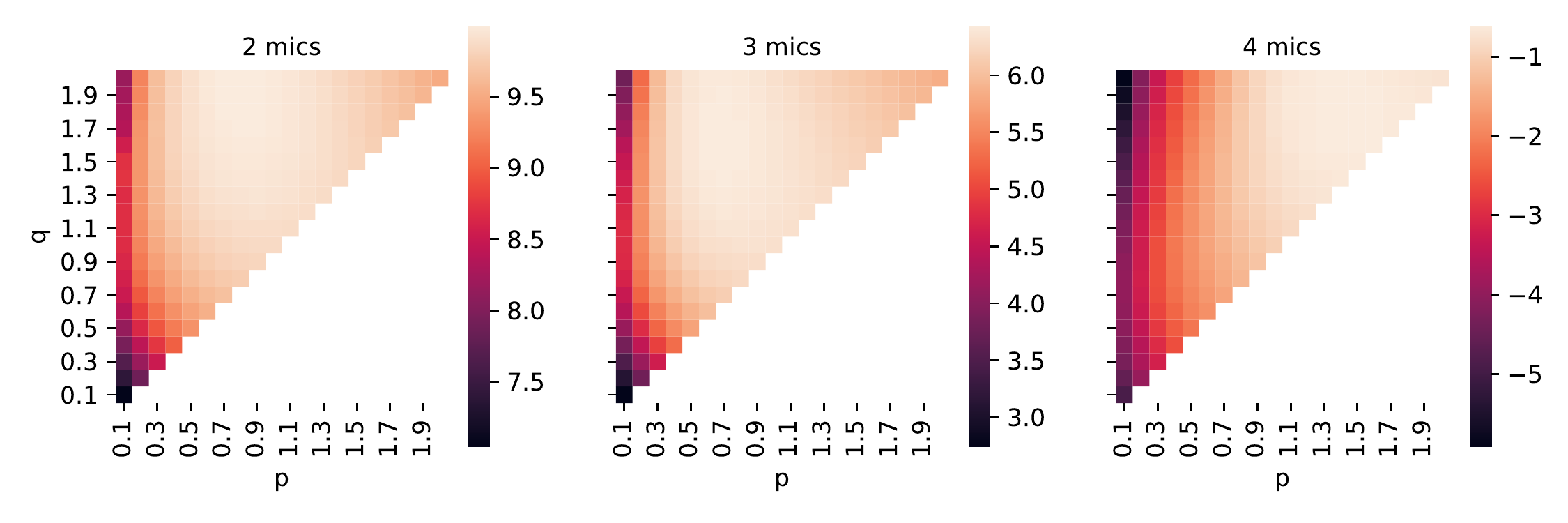}
    \caption{AuxIVA, SDR}
    \flabel{auxiva_sdr}
  \end{subfigure}
  \begin{subfigure}{.5\textwidth}
    \centering
    \includegraphics[width=\linewidth]{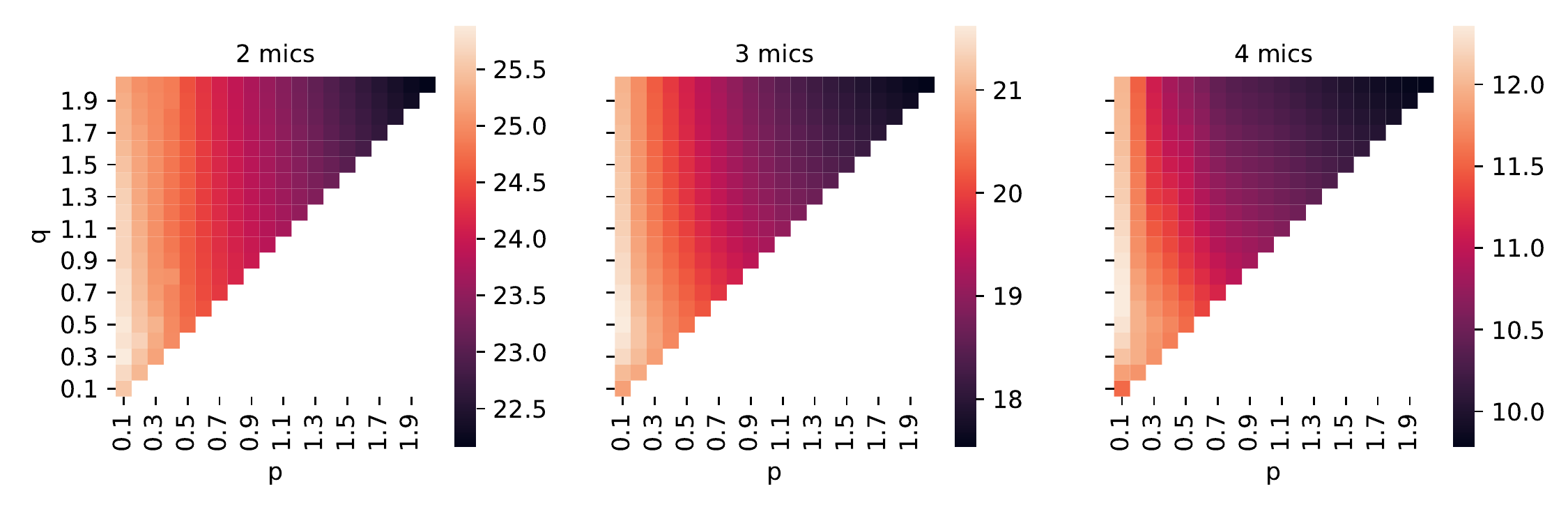}
    \caption{AuxIVA, SIR}
    \flabel{auxiva_sir}
  \end{subfigure}
  \begin{subfigure}{.5\textwidth}
    \centering
    \includegraphics[width=\linewidth]{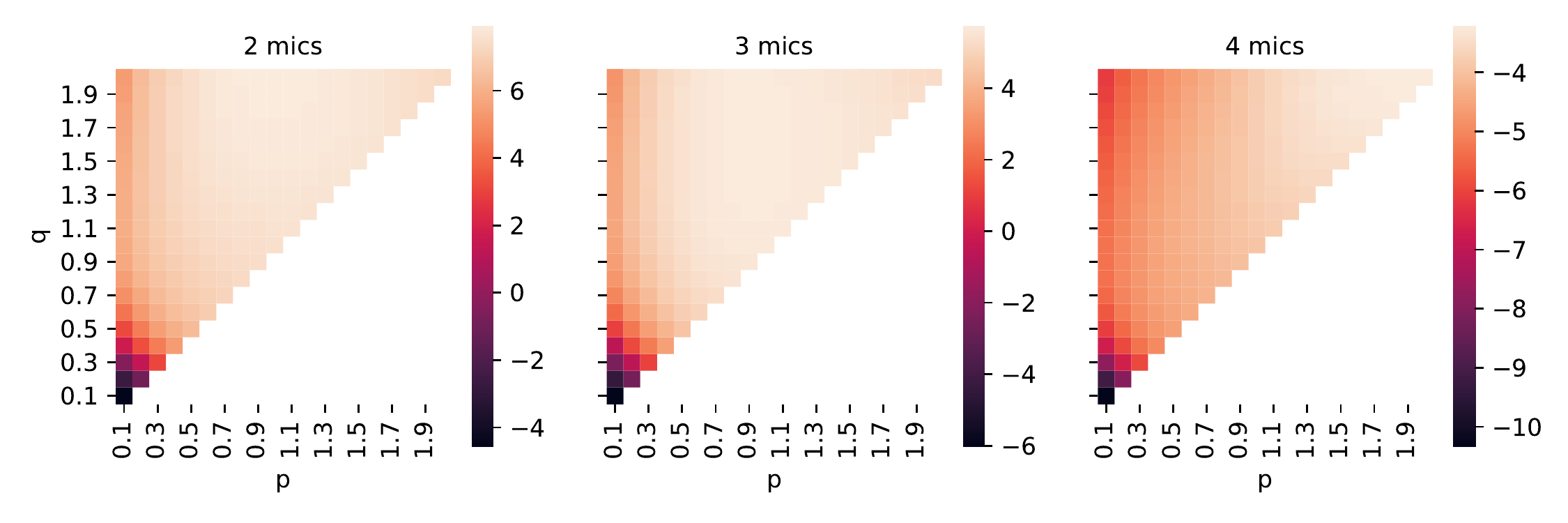}
    \caption{ILRMA, SDR}
    \flabel{ilrma_sdr}
  \end{subfigure}
  \begin{subfigure}{.5\textwidth}
    \centering
    \includegraphics[width=\linewidth]{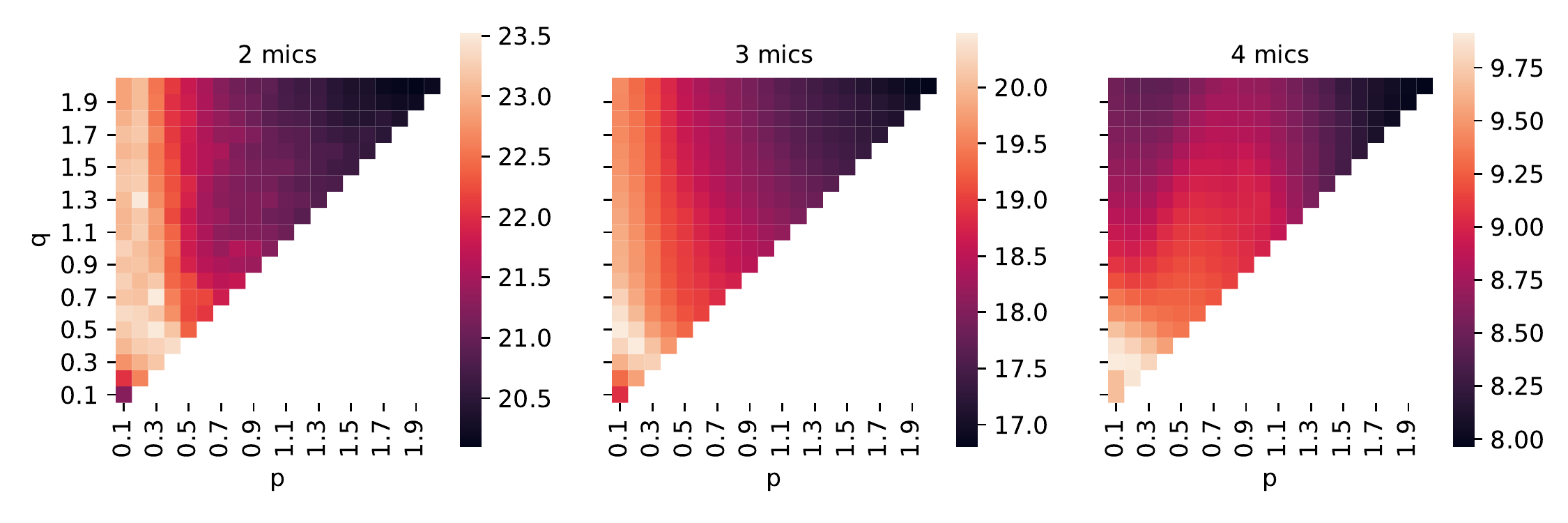}
    \caption{ILRMA, SIR}
    \flabel{ilrma_sir}
  \end{subfigure}
  \begin{subfigure}{.5\textwidth}
    \centering
    \includegraphics[width=\linewidth]{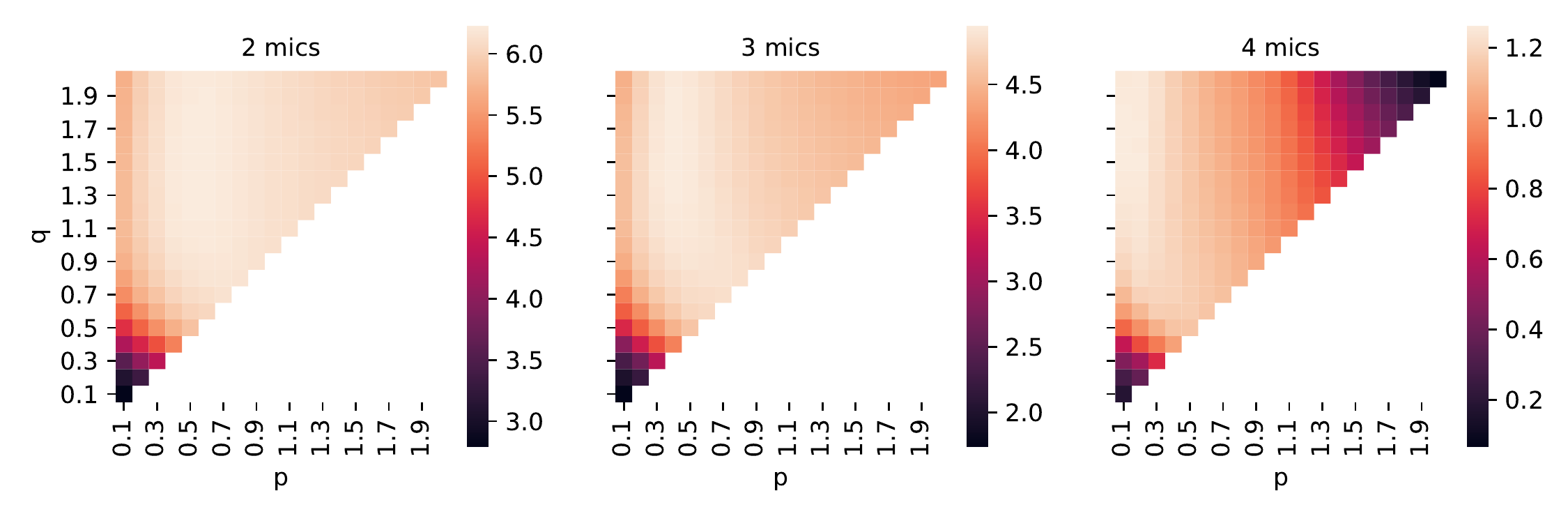}
    \caption{ILRMA-T, SDR}
    \flabel{ilrma_t_sdr}
  \end{subfigure}
  \begin{subfigure}{.5\textwidth}
    \centering
    \includegraphics[width=\linewidth]{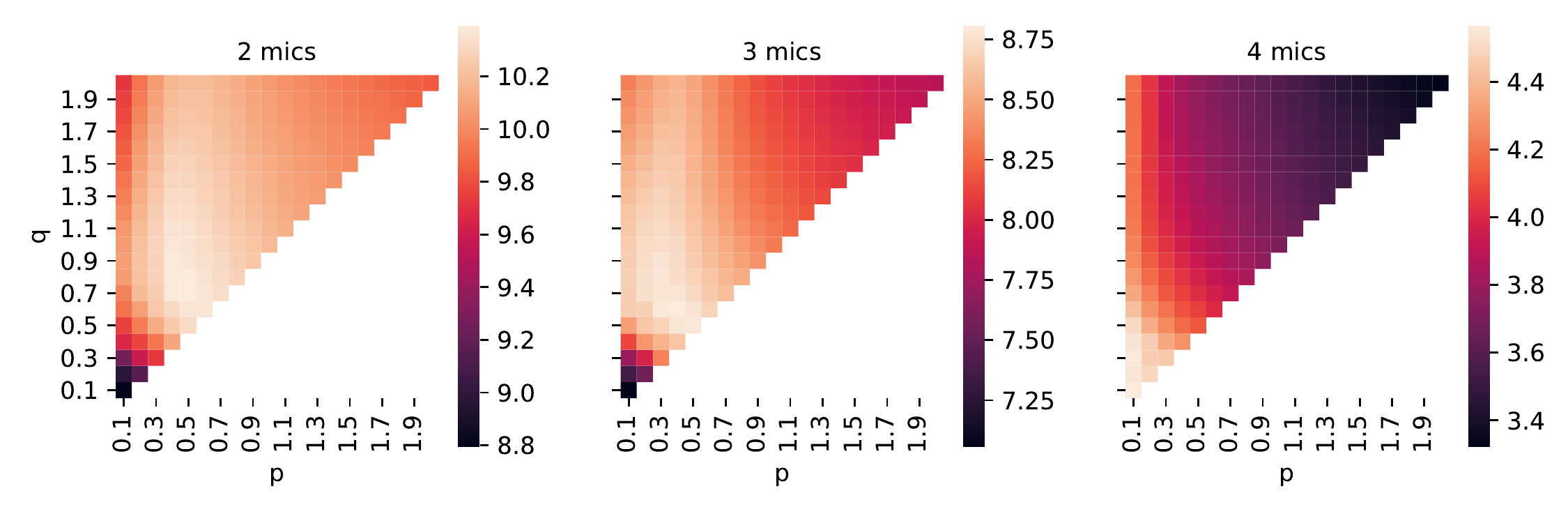}
    \caption{ILRMA-T, SIR}
    \flabel{ilrma_t_sir}
  \end{subfigure}
  \caption{Average SDR and SIR performance of GMDP. Bright and dark colors indicate high and low performance, respectively.}
  \flabel{param_sweep}
\end{figure*}

Consider the residual signal, where the scaling factor $a_{mkf}$ is assumed known such that $h_{mkf} s_{kfn} = a_{mkf} y_{kfn}$, then
\begin{equation}
  e_{mfn} = x_{mfn} - a_{mkf} y_{kfn} = \sum_{\ell \neq k} h_{mkf} s_{\ell fn} + b_{mfn}.
  \elabel{residual_model}
\end{equation}
In other words, for the correct scaling factor, the residual is identical to the mixing model \eref{mix} with the $k$th source removed.
Note that sources are typically assumed to have non-Gaussian distributions.
If their number is large, the central limit theorem may be invoked to justify Gaussianity of $e_{mfn}$.
However, the typical number of sources in practice is small, e.g. two to four.
In this case, the residual is also expected to be non-Gaussian.

We propose a generalized minimal distortion principle (GMDP) with a tunable error model.
To allow modelling of inter-frequency and inter-frame dependencies, we consider the residual spectrograms
\begin{equation}
  \mE_{mk}(\vz) = \mX_m - \diag(\vz) \mY_k,\quad m=1,\ldots,M,
\end{equation}
where $(\mX_m)_{fn} = x_{mfn}$, $(\mY_k)_{fn} = y_{kfn}$, and $\vz = [z_1,\ldots,z_F]^\top$.
Then, the MLE of $\vz_{mk}$ is
\begin{equation}
  \{ \hat{\vz}_{mk} \}_{m=1}^M = \underset{\vz_1,\ldots,\vz_M\in\C^F}{\arg\min}\ -\log p_e\left(\{\mE_{mk}(\vz_m)\}_{m=1}^M\right),
  \nonumber
\end{equation}
where $p_e$ is the probability density function corresponding to the chosen error model.
Note that $\hat{\vz}_{mk} = [a_{mk1},\ldots,a_{mkF}]^\top$.

While there are lots of possible choices for the error function, inspired by the success of spherical contrast functions in IVA~\cite{Ono:2011tn}, we propose to use $p_e$ such that
\begin{equation}
  -\log p_e(\{ \mE_m\}_{m=1}^M) = \sum_{m=1}^M \| \mE_m \|_{p,q}^p + \text{const},
\end{equation}
where the mixed $\ell_{p,q}$-norm is defined as
\begin{equation}
  \|\mathbf{E}\|_{p,q} = \left( \sum_n\left(  \sum_f |e_{fn}|^q \right)^{\frac{p}{q}} \right)^{\frac{1}{p}}.
  \elabel{mixed_norm}
\end{equation}
In Laplace AuxIVA~\cite{Ono:2011tn}, an $\ell_{1,2}$-norm is used with the effect of promoting group sparsity in the frames.
That is, active frames are sparse, but within an active frame, frequencies are not sparse.
Using an $\ell_{p,q}$-norm, it is possible to optimize the desired sparsity of both frames, and frequency components.
This makes sense for speech and music signals that are typically somewhat sparse in both due to harmonics and non-stationarity.

\section{Optimization with MM Algorithm}

\begin{table*}
  \centering
  \caption{Average performance of different algorithms. SDR/SIR refer to SI-SDR/SI-SIR~\cite{LeRoux:2018tq} for AuxIVA and ILRMA, and to \textit{bss\_eval}'s SDR/SIR~\cite{Vincent:2006fz} for ILRMA-T. This is due to the latter using the anechoic signals as reference.}
  \begin{tabular}{@{}l@{~}rrrrrrrrrrrrr@{}}
    \toprule
   & & \multicolumn{2}{c}{PB} & \multicolumn{2}{c}{MDP} & \multicolumn{2}{c}{GMDP/SDR} & \multicolumn{2}{c}{GMDP/SIR} & \multicolumn{2}{c}{GMDP/SIR-10} & \multicolumn{2}{c}{GMDP/SDR-F} \\
   \cmidrule{3-14}
  Algo. & Mics  &  SDR  &  SIR  &  SDR  &  SIR  &  SDR  &  SIR  &  SDR  &  SIR  &  SDR  &  SIR  &  SDR  &  SIR  \\
    \midrule
    AuxIVA~\cite{Ono:2011tn} & 2  & 9.72 & 22.78 & 9.50 & 22.16 & \textbf{9.99} & 23.94 & 9.51 & \textbf{25.04} & 9.51 & \textbf{25.04} & 9.95 & 23.49 \\
                             & 3  & 6.01 & 17.92 & 5.85 & 17.53 & \textbf{6.43} & 19.34 & 5.92 & \textbf{20.55} & 5.92 & \textbf{20.55} & 6.33 & 18.63 \\
                             & 4  & -0.71 & 9.96 & -0.72 & 9.78 & \textbf{-0.62} & 10.14 & -0.72 & \textbf{10.45} & -0.72 & \textbf{10.45} & -0.68 & 10.36 \\
    ILRMA~\cite{Kitamura:2016vj} & 2  & 7.06 & 21.55 & 7.36 & 20.19 & \textbf{7.91} & 20.94 & 7.46 & \textbf{21.96} & 7.46 & \textbf{21.96} & 7.84 & 20.63 \\
                                 & 3  & 5.14 & 17.91 & 5.25 & 16.80 & \textbf{5.74} & 18.08 & 5.27 & \textbf{18.99} & 5.27 & \textbf{18.99} & 5.63 & 17.49 \\
                                 & 4  & -4.05 & 8.44 & -3.24 & 7.96 & \textbf{-3.22} & 8.06 & -3.22 & 8.11 & -3.22 & 8.11 & -3.41 & \textbf{8.48} \\
    \midrule
    ILRMA-T~\cite{Ikeshita:2019bl} & 2  & 5.99 & 10.02 & 5.87 & 9.84 & \textbf{6.23} & 10.26 & 6.09 & \textbf{10.39} & 6.14 & 10.38 & 6.21 & 10.28 \\
                                   & 3  & 4.44 & 8.00 & 4.35 & 7.84 & \textbf{4.94} & 8.62 & 4.66 & \textbf{8.81} & 4.81 & 8.77 & \textbf{4.94} & 8.64 \\
                                   & 4  & 0.32 & 3.37 & 0.07 & 3.32 & \textbf{1.26} & 4.19 & 0.19 & \textbf{4.56} & 1.23 & 4.22 & 1.18 & 3.88 \\
    \bottomrule
  \end{tabular}
  \tlabel{performance}
\end{table*}

The scaling factor to obtain the source images under the proposed GMDP are
\begin{equation}
  \hat{\vz}_{mk} = \underset{\mathbf{z} \in \mathbb{C}^{F}}{\arg\min}\ \| \mE_{mk}(\vz) \|_{p,q}^p.
\end{equation}
Unlike the MDP, there is unfortunately no closed form solution for this problem.
Nevertheless, it can be efficiently tackled by an IRLS scheme derived using the MM technique~\cite{Lange:2016wp}.

For the optimization of an objective $f(\vtheta)$, the MM technique introduces a surrogate $Q(\vtheta, \hat{\vtheta})$ such that
\begin{align}
  Q(\hat{\vtheta}, \hat{\vtheta}) = f(\hat{\vtheta}),\ \text{and}\  Q(\vtheta, \hat{\vtheta}) \geq f(\vtheta),    \quad \forall \vtheta,\hat{\vtheta}.
  \elabel{surrogate_ineq}
\end{align}
Then, the sequence of iterates
\begin{align}
  \vtheta_{t} = \underset{\vtheta}{\arg\min}\ Q(\vtheta, \vtheta_{t-1}), \quad t=1,\ldots, T,
\end{align}
monotonically decreases the cost function since,
\begin{equation}
  f(\vtheta_{t-1}) = Q(\vtheta_{t-1}, \vtheta_{t-1}) \geq \underset{\vtheta}{\min}\ Q(\vtheta, \vtheta_{t-1}) \geq f(\vtheta_t).
  \nonumber
\end{equation}

To construct the surrogate, we use the following inequality,
\begin{equation}
r^q \leq \frac{q}{2 r_0^{2 - q}}r^2 + \text{const}, \quad 0< q \leq 2,
\elabel{inequality}
\end{equation}
that is derived from an inequality for super-Gaussian sources~\cite{Ono:2010hh} or from concave-convex arguments~\cite{Murata:2018ft}.
For values of $0 < p \leq q \leq 2$, by applying~\eref{inequality} twice, we have
\begin{align}
  Q(\mE\;;\; \wh{\mE}) = \sum_{n,f} w_{fn}(\wh{\mE})  |e_{fn}|^2 + C \geq \| \mE \|_{p,q}^p,
\end{align}
where $C$ is a constant and with weights
\begin{equation}
  w_{fn}(\wh{\mE}) = p\left(2 \left(\sum\limits_{f^\prime=1}^F |\hat{e}_{f^\prime n}|^q \right)^{\left(1 - \frac{p}{q}\right)} |\hat{e}_{fn}|^{2 - q}\right)^{-1}.
\end{equation}
Equality holds for $\mE = \wh{\mE}$.
%

Finally, given the current iterate $\vz_t$, the next iterate is obtained by minimizing $Q(\mE_{mk}(\vz)\;;\;\mE_{mk}(\vz_t))$, i.e.,
\begin{equation}
\mathbf{z}_{t+1} \gets \frac{\sum_n w_{fn}(\mathbf{z}_t) x_{fn} y^*_{fn}}{\sum_n w_{fn}(\mathbf{z}_t) |y_{fn}|^2}
\end{equation}
where $w_{fn}(\mathbf{z})$ is short for $w_{fn}(\mE_{mk}(\vz))$.

\section{Experiments}
\seclabel{experiments}

\begin{table}
  \centering
  \caption{Best parameters for the proposed algorithm.}
  \resizebox{\linewidth}{!}{%
  \begin{tabular}{@{}l@{}rrrrrrrrrr@{}}
    \toprule
    \multicolumn{2}{@{}l@{}}{\textbf{Criteria}}  & \multicolumn{3}{c}{ SDR }  & \multicolumn{3}{c}{ SIR-$10$ } & \multicolumn{3}{c}{ SDR-F} \\
    \cmidrule{1-2}\cmidrule{3-5} \cmidrule{6-8} \cmidrule{9-11}
    \text{Algo.} & \text{Mics}  & $p$ & $q$ & $N$ & $p$ & $q$ & $N$ & $p$ & $q$ & $N$ \\
    \midrule
    AuxIVA & 2  & 0.8 & 1.9 & 4 & 0.4 & 0.8 & 9 & 1.1 & 1.7 & 4 \\
           & 3  & 0.7 & 1.6 & 5 & 0.3 & 1.0 & 8 & 1.1 & 1.7 & 4 \\
           & 4  & 1.4 & 1.8 & 3 & 1.1 & 1.5 & 4 & 1.1 & 1.7 & 4 \\
    ILRMA & 2  & 1.0 & 2.0 & 3 & 0.5 & 1.4 & 6 & 1.3 & 1.9 & 3 \\
          & 3  & 0.9 & 1.6 & 4 & 0.5 & 0.9 & 8 & 1.3 & 1.9 & 3 \\
          & 4  & 1.8 & 2.0 & 3 & 1.7 & 2.0 & 3 & 1.3 & 1.9 & 3 \\
    ILRMA-T & 2  & 0.6 & 1.5 & 5 & 0.5 & 0.8 & 10 & 0.4 & 1.5 & 6 \\
            & 3  & 0.4 & 1.6 & 6 & 0.3 & 0.9 & 10 & 0.4 & 1.5 & 6 \\
            & 4  & 0.1 & 1.7 & 8 & 0.1 & 1.1 & 10 & 0.4 & 1.5 & 6 \\
    \bottomrule
  \end{tabular}
}
\tlabel{opt_params}
\end{table}

\subsection{Setup}

We use the \texttt{pyroomacoustics} package to simulate a hundred random rooms with reverberation time ($T_{60}$) between \SI{60}{\milli\second} and \SI{500}{\milli\second}~\cite{Scheibler:2018di}.
The microphone array is circular with diameter chosen so that neighboring elements are \SI{2}{\centi\meter} apart, and is placed at random in the room.
The sources are placed at random but so that they fall within $[d_{\text{crit}}, d_{\text{crit}} + 1\,\si{\meter}]$ from the array.
Here, $d_{\text{crit}} = 0.057\sqrt{V/T_{60}}\,\si{\meter}$ is the critical distance, with $V$ being the volume of the room~\cite{Kuttruff:2009uq}.
The source signals are speech utterances from the CMU Arctic database~\cite{Kominek:2004vf,cmu_conc_15}.

We evaluate the effectiveness of GMDP on three determined BSS algorithms (i.e.~same number of sources and microphones):
AuxIVA~\cite{Ono:2011tn}, IVA with simple power based source model,
ILRMA~\cite{Kitamura:2016vj}, IVA with non-negative low-rank model
ILRMA-T~\cite{Ikeshita:2019bl}, joint BSS and dereverberation algorithm with non-negative low-rank source model.
Then, the source image are further estimated with PB~\cite{Murata:2001gb}, MDP~\cite{Matsuoka:2001da}, and the proposed GMDP.
For GMDP, the values of $p$ and $q$ are further sweeped in increments of 0.1 in their range so that we can find the best parameters.
AuxIVA and ILRMA are evaluated in terms of scale invariant signal-to-distortion ratio (SI-SDR) and signal-to-interference ratio (SI-SIR)~\cite{LeRoux:2018tq} and the clean reverberant microphone signals are used as reference.
The evaluation of ILRMA-T is different since it also dereverberates the signals.
We evaluate it with \texttt{bss\_eval}~\cite{Vincent:2006fz} in terms of conventional SDR and SIR, on the clean, anechoic microphone signals.
The metric from \texttt{bss\_eval} forgives a 512 taps filter (\SI{32}{\milli\second} at \SI{16}{\kilo\hertz}), which accounts for the residual reverberation.
For GMDP, we run the MM algorithm for 100 iterations or until $\| \vz_t - \vz_{t-1} \|/\|\vz_{t-1}\| \leq 0.01$, whichever comes first.
The reference to compute the source images is the first microphone.

\subsection{Result}

\ffref{param_sweep} shows the result of the parameter sweep for $p$ and $q$.
The heatmaps show high and low SDR/SIR with brighter and darker colors, respectively.
We observe that general trends of SDR and SIR are quite different.
SDR is poor for small values of $p,q$ and generally improves going towards 2.
On the contrary, SIR improves towards smaller values.
This is expected since SDR measures faithfulness to the reference signal while SIR measures effectiveness of separation.

In \tref{performance}, we compare the performance of PB and MDP to that of GMDP.
Since a balance between SDR and SIR needs to be found, we compare a few strategies for picking the best $p$ and $q$.
We first note that the proposed method performs better for all strategies.
GMDP/SDR chooses $p$ and $q$ yielding the highest SDR.
Under this choice, the SIR gain tends to be modest.
However, from \ffref{param_sweep} we also note that the SDR changes little over a large range of parameter values.
Thus, the strategy GMDP/SIR chooses $p$ and $q$ yielding the largest SIR under the constraint that the SDR is no less than that of MDP (i.e., $p=q=2$).
In this case, the SIR increases by about \SI{1}{\decibel} to \SI{2}{\decibel} for most algorithms with no decrease of SDR.
Now, in some cases, especially for ILRMA-T, some parameters may lead to a large iteration count of the MM algorithm.
The GMDP/SIR-10 strategy is the same as the previous one, but further limits the median iteration count to 10.
This is achieved at very little cost in either metrics.
Finally, it may be of practical interest to fix $p$ and $q$ independent of the channel count.
GMDP/SDR-F maximizes the average SDR over all channel counts.
While still improving over PB and MDP in most cases, the gain is more modest and case-by-case choice of $p$ and $q$ seems necessary to obtain the best performance.
This makes sense since the spectrogram error distribution varies according to the number of residual sources.
\tref{opt_params} contains all the parameters used in this experiment and the median number of iterations of the MM algorithm.

\section{Conclusions}
\seclabel{conclusions}

We proposed a new method for the estimation of source images from the signals separated by BSS algorithms.
The method generalizes the traditional minimal distortion principle to maximum likelihood estimation with a problem specific residual spectrogram model.
Concretely, we proposed to minimize a mixed-norm that allows to promote sparsity at different rates in time and frequency.
The optimization is carried out by a simple MM algorithm that is both fast and straightforward to implement.
We demonstrate the effectiveness of the method on several BSS and joint BSS-dereverberation algorithms.
We show that the proposed method allows to improve the separation by \SI{1}{\decibel} to \SI{2}{\decibel} without degradation in SDR and with minimal computational overhead.
Finally, we point out that the proposed method can be combined with further post-processing using beamforming, as recently proposed~\cite{Araki:2019gn}.

\bibliographystyle{IEEEtran}
\bibliography{IEEEabrv,refs}

\begin{thebibliography}{10}
\providecommand{\url}[1]{#1}
\csname url@samestyle\endcsname
\providecommand{\newblock}{\relax}
\providecommand{\bibinfo}[2]{#2}
\providecommand{\BIBentrySTDinterwordspacing}{\spaceskip=0pt\relax}
\providecommand{\BIBentryALTinterwordstretchfactor}{4}
\providecommand{\BIBentryALTinterwordspacing}{\spaceskip=\fontdimen2\font plus
\BIBentryALTinterwordstretchfactor\fontdimen3\font minus
  \fontdimen4\font\relax}
\providecommand{\BIBforeignlanguage}[2]{{%
\expandafter\ifx\csname l@#1\endcsname\relax
\typeout{** WARNING: IEEEtran.bst: No hyphenation pattern has been}%
\typeout{** loaded for the language `#1'. Using the pattern for}%
\typeout{** the default language instead.}%
\else
\language=\csname l@#1\endcsname
\fi
#2}}
\providecommand{\BIBdecl}{\relax}
\BIBdecl

\bibitem{Comon:1512057}
P.~Comon and C.~Jutten, \emph{Handbook of blind source separation: independent
  component analysis and applications}, 1st~ed.\hskip 1em plus 0.5em minus
  0.4em\relax Oxford, UK: Academic Press/Elsevier, 2010.

\bibitem{Comon:1994kr}
P.~Comon, ``Independent component analysis, a new concept?'' \emph{Signal
  Processing}, vol.~36, no.~3, pp. 287--314, 1994.

\bibitem{Smaragdis:1998kl}
P.~Smaragdis, ``Blind separation of convolved mixtures in the frequency
  domain,'' \emph{Neurocomputing}, vol.~22, no. 1-3, pp. 21--34, Nov. 1998.

\bibitem{Sawada:fk}
H.~Sawada, S.~Araki, and S.~Makino, ``Measuring dependence of bin-wise
  separated signals for permutation alignment in frequency-domain {BSS},'' in
  \emph{Proc. IEEE ISCAS}, New Orleans, LA, USA, May 2007, pp. 3247--3250.

\bibitem{Kim:2006ex}
T.~Kim, H.~T. Attias, S.-Y. Lee, and T.-W. Lee, ``Blind source separation
  exploiting higher-order frequency dependencies,'' \emph{IEEE Trans. Audio,
  Speech, Language Process.}, vol.~15, no.~1, pp. 70--79, Dec. 2006.

\bibitem{Hiroe:2006ib}
A.~Hiroe, ``Solution of permutation problem in frequency domain {ICA}, using
  multivariate probability density functions,'' in \emph{ASIACRYPT 2016}.\hskip
  1em plus 0.5em minus 0.4em\relax Berlin, Heidelberg: Springer Berlin
  Heidelberg, 2006, pp. 601--608.

\bibitem{Murata:2001gb}
N.~Murata, S.~Ikeda, and A.~Ziehe, ``An approach to blind source separation
  based on temporal structure of speech signals,'' \emph{Neurocomputing},
  vol.~41, no. 1-4, pp. 1--24, Oct. 2001.

\bibitem{Koldovsky:2017ez}
Z.~Koldovsk{\'{y}} and F.~Nesta, ``Performance analysis of source image
  estimators in blind source separation,'' \emph{{IEEE} Trans. Signal
  Process.}, vol.~65, no.~16, pp. 4166--4176, Jun. 2017.

\bibitem{Tibshirani:1996dm}
R.~Tibshirani, ``Regression shrinkage and selection via the lasso,'' \emph{J R
  STAT SOC B}, vol.~58, no.~1, pp. 267--288, Jan. 1996.

\bibitem{Candes:2008hb}
E.~J. Candes and M.~B. Wakin, ``An introduction to compressive sampling,''
  \emph{{IEEE} Signal Process. Mag.}, vol.~25, no.~2, pp. 21--30, Mar. 2008.

\bibitem{Lange:2016wp}
K.~Lange, \emph{{MM} optimization algorithms}.\hskip 1em plus 0.5em minus
  0.4em\relax SIAM, 2016.

\bibitem{Murata:2018ft}
N.~Murata, S.~Koyama, N.~Takamune, and H.~Saruwatari, ``Sparse representation
  using multidimensional mixed-norm penalty with application to sound field
  decomposition,'' \emph{{IEEE} Trans. Signal Process.}, vol.~66, no.~12, pp.
  3327--3338, May 2018.

\bibitem{Daubechies:2010hf}
I.~Daubechies, R.~DeVore, M.~Fornasier, and C.~S. G{\"u}nt{\"u}rk,
  ``Iteratively reweighted least squares minimization for sparse recovery,''
  \emph{Communications on Pure and Applied Mathematics}, vol.~63, no.~1, pp.
  1--38, Jan. 2010.

\bibitem{Kowalski:2009ge}
M.~Kowalski, ``Sparse regression using mixed norms,'' \emph{Appl Comput Harmon
  A}, vol.~27, no.~3, pp. 303--324, Nov. 2009.

\bibitem{Ono:2011tn}
N.~Ono, ``Stable and fast update rules for independent vector analysis based on
  auxiliary function technique,'' in \emph{Proc. {IEEE} WASPAA}, New Paltz, NY,
  USA, Oct. 2011, pp. 189--192.

\bibitem{Kitamura:2016vj}
D.~Kitamura, N.~Ono, H.~Sawada, H.~Kameoka, and H.~Saruwatari, ``Determined
  blind source separation unifying independent vector analysis and nonnegative
  matrix factorization,'' \emph{{IEEE/ACM} Trans. Audio Speech Lang. Process.},
  2016.

\bibitem{Ikeshita:2019bl}
R.~Ikeshita, N.~Ito, T.~Nakatani, and H.~Sawada, ``A unifying framework for
  blind source separation based on a joint diagonalizability constraint,'' in
  \emph{Proc. {IEEE} EUSIPCO}, Sep. 2019.

\bibitem{Allen:1977in}
J.~Allen, ``Short term spectral analysis, synthesis, and modification by
  discrete {F}ourier transform,'' \emph{IEEE Trans. Acoust., Speech, Signal
  Process.}, vol.~25, no.~3, pp. 235--238, Jun. 1977.

\bibitem{Hyvarinen:ek}
A.~Hyv\"{a}rinen, ``Fast and robust fixed-point algorithms for independent
  component analysis,'' \emph{{IEEE} Trans. Neural Netw.}, vol.~10, no.~3, pp.
  626--634, May 1999.

\bibitem{Matsuoka:2001da}
K.~Matsuoka and S.~Nakashima, ``Minimal distortion principle for blind source
  separation,'' in \emph{Proc. ICA}, San Diego, Dec. 2001, pp. 722--727.

\bibitem{Matsuoka:2002jy}
K.~Matsuoka, ``Minimal distortion principle for blind source separation,'' in
  \emph{Proc. SICE}, Osaka, Japan, Aug. 2002, pp. 2138--2143.

\bibitem{LeRoux:2018tq}
J.~Le~Roux, S.~Wisdom, H.~Erdogan, and J.~R. Hershey, ``{SDR} - half-baked or
  well done?'' in \emph{Proc. {IEEE} ICASSP}, Brighton, UK, May 2019, pp.
  626--630.

\bibitem{Vincent:2006fz}
E.~Vincent, R.~Gribonval, and C.~Fevotte, ``Performance measurement in blind
  audio source separation,'' \emph{IEEE Trans. Audio, Speech, Language
  Process.}, vol.~14, no.~4, pp. 1462--1469, Jun. 2006.

\bibitem{Ono:2010hh}
N.~Ono and S.~Miyabe, ``Auxiliary-function-based independent component analysis
  for super-{G}aussian sources,'' \emph{Proc. LVA/ICA}, vol. 6365, no.~6, pp.
  165--172, Sep. 2010.

\bibitem{Scheibler:2018di}
R.~Scheibler, E.~Bezzam, and I.~Dokmani{\'c}, ``Pyroomacoustics: A {P}ython
  package for audio room simulations and array processing algorithms,'' in
  \emph{Proc. {IEEE} ICASSP}, Calgary, CA, Apr. 2018, pp. 351--355.

\bibitem{Kuttruff:2009uq}
H.~Kuttruff, \emph{Room acoustics}.\hskip 1em plus 0.5em minus 0.4em\relax CRC
  Press, 2009.

\bibitem{Kominek:2004vf}
J.~Kominek and A.~W. Black, ``{CMU ARCTIC} databases for speech synthesis,''
  Language Technologies Institute, School of Computer Science, Carnegie Mellon
  University, Tech. Rep. CMU-LTI-03-177, 2003.

\bibitem{cmu_conc_15}
\BIBentryALTinterwordspacing
R.~Scheibler, ``{CMU ARCTIC} concatenated 15s,'' {Z}enodo. [Online]. Available:
  \url{http://doi.org/10.5281/zenodo.3066489}
\BIBentrySTDinterwordspacing

\bibitem{Araki:2019gn}
S.~Araki, N.~Ono, K.~Kinoshita, and M.~Delcroix, ``Projection back onto
  filtered observations for speech separation with distributed microphone
  array,'' in \emph{Proc. {IEEE} CAMSAP}, Le Gosier, Guadeloupe, Guadeloupe,
  Dec. 2019, pp. 291--295.

\end{thebibliography}

\end{document}